\documentclass[preprint,5p,times]{elsarticle}
\usepackage{lineno}
\usepackage[T1]{fontenc}
\usepackage[utf8]{inputenc}
\usepackage{graphicx}
\usepackage{textcomp}
\usepackage{units}
\usepackage{amssymb,amsmath}

\journal{NIM A}

\begin{document}
\begin{frontmatter}
\title{The multi-purpose three-axis spectrometer (TAS) MIRA at FRM II}
\author[frm,e21]{R. Georgii}
\ead{Robert.Georgii@frm2.tum.de}
\author[frm,e21]{T. Weber}
\author[frm,e21,jcns]{G. Brandl}
\author[frm,e21]{M. Skoulatos}
\author[LA]{M. Janoschek}
\author[frm]{S. M\"uhlbauer}
\author[e21]{C. Pfleiderer}
\author[e21]{P. B\"oni}
\address[frm]{Heinz Maier-Leibnitz Zentrum (MLZ),
  Technische Universität München, Lichtenbergstr.\ 1, 85747 Garching, Germany}
\address[e21]{Physik Department E21,  Technische Universität München,
  James-Franck-Str., 85747 Garching, Germany}
 \address[jcns]{Heinz Maier-Leibnitz Zentrum (MLZ),  J\"ulich Center for Neutron Scattering (JCNS),
  Lichtenbergstr., 85747 Garching, Germany}
 \address[LA]{Condensed Matter and Magnet Science Group, Los Alamos National Laboratory, Los Alamos, NM 87545, USA}

\date{\today}

\begin{abstract}
The cold-neutron three-axis spectrometer MIRA is an instrument optimized for low-energy excitations. Its excellent intrinsic $Q$-resolution makes it ideal for studying incommensurate magnetic systems (elastic and inelastic). MIRA is at the forefront of using advanced neutron focusing optics such as elliptic guides, which enable the investigation of small samples under extreme conditions. Another advantage of MIRA is the modular assembly allowing for instrumental adaption to the needs of the experiment within a few hours. The development of new methods such as the spin-echo technique MIEZE is another important application at MIRA. Scientific topics include the investigation of complex inter-metallic alloys and spectroscopy on incommensurate magnetic structures.
\end{abstract}

\end{frontmatter}

\section{Instrument description}
MIRA is a multipurpose instrument that has been in operation at the FRM II since 2004. It utilizes neutrons from the cold source, fed via the $m = 2.0$ neutron guide NL6  to the two beamports MIRA-1 and MIRA-2. The lower cut-off wavelength of the neutron guide NL6 is 2.0~\AA. In the three-axis mode the major advantages of the instrument are the excellent resolution in momentum transfer, $Q$, the low background, as well as the  option to investigate small samples by means of focusing guides before and after the sample position.
\begin{figure}[htb]
  \begin{center}
   \includegraphics[width=55mm]{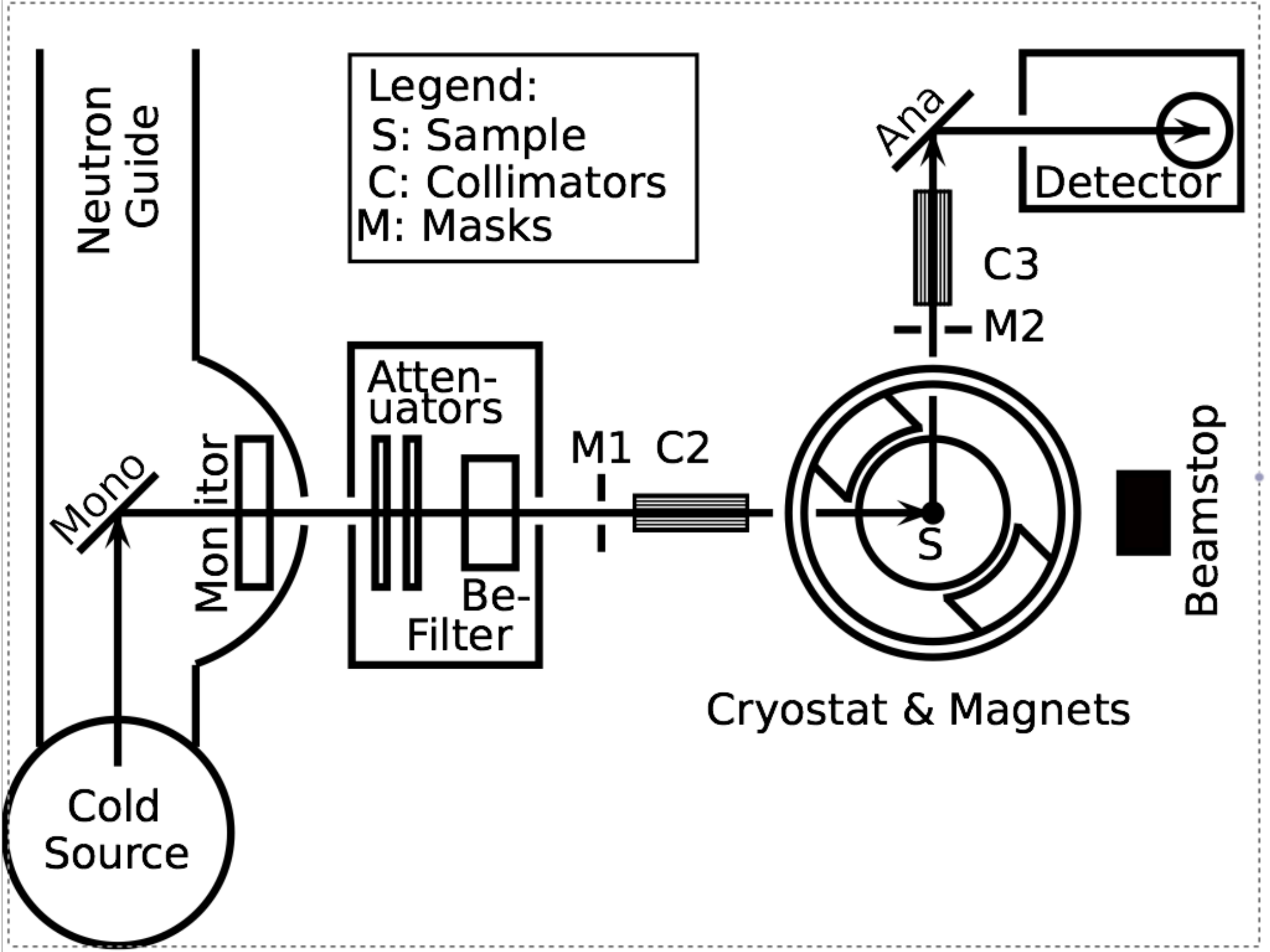}
   \includegraphics[width=55mm]{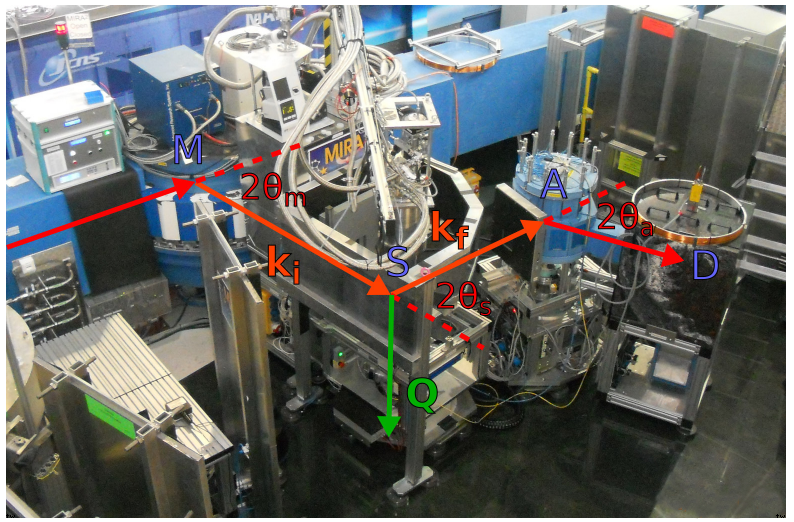}
   \caption{\label{fig:schema} A schematic view (top) and a photo \cite{Tobi_PhD} (bottom) of MIRA-2 in TAS mode. In the photo the position of the graphite monochromator (M), the sample position (S), the graphite analyzer (A) and the detector (D) are marked. The in-coming neutron beam ($k_i$), the out-going beam ($k_f$), the monochromotor take-off angle ($2 \theta_m$), the scattering angle ($\theta_S$) and the direction of the momentum transfer $Q$ are sketched for clarity.} 
\end{center}
\end{figure}

There are several instrument configurations which can be changed within a short time. The control software NICOS is adapted to the flexibility of the instrument and supports such quick changes of the operational mode of MIRA. The $Q$-transfer can be varied in the range of $10^{-3}$~\AA$^{-1} \le Q \le 2.5$~\AA$^{-1}$.  We note that the lowest  Q is nearly comparable to what can be achieved by means of small angle neutron scattering (SANS) highlighting the excellent Q-resolution. 

The two beam ports allow the extraction of neutrons in a range of wavelengths 3.5~\AA\ $\le \lambda \le$ 20~\AA. MIRA-2 provides shorter wavelengths from an array of seven horizontally focusing HOPG monochromators with a variable focus length delivering an intensity of $1.1\cdot 10^7$s$^{-1}$at 4.75~\AA~for a typical horizontal divergence of 80' and vertical of 40', respectively. The higher-order wavelength contributions below 4.2~\AA\ are filtered out with a cooled Be-filter. Very cold neutrons with a wavelength of 9~\AA\ are provided at MIRA-1 under a take-off angle of $90^\circ$ by an intercalated HOPG monochromator yielding an intensity of $10^6$s$^{-1}$. Here the divergence is significantly larger due to the longer wavelengths, i.e. $\alpha = 2^\circ$. Using a multilayer monochromator instead, wavelengths $\lambda > 10~\AA$ can be used. However, due to the reflecting properties of the multilayer monochromator a small contamination with longer wavelengths is unavoidable. 

The instrument is fully compatible with all sample environment of the FRM II (dry cryostats, furnaces, magnets). Sample environments from other instruments, such as an Eulerian cradle may be installed on demand. In addition, MIRA offers electro magnets for vertical and horizontal fields up to 0.3~T (in and out of plane) and also a dedicated closed-cycle cryostat for sample temperatures ranging from room temperature down to 3.5~K. 

For adapting the instrument to the scientific needs of the users the different setups of the instrument comprise:
\subparagraph*{The standard three-axis spectroscopy (TAS) mode} for inelastic scattering experiments with an excellent $Q$-resolution (Fig.~\ref{fig:schema}), i.e.\ a typical full-width at half-maximum of $\Delta Q = 0.014$~\AA$^{-1}$ at $Q = 2$~\AA$^{-1}$ and an incident neutron energy of $E_i = 4$~meV. The high $Q$-resolution is achieved mainly by a relatively large distance of 115 cm between sample and analyzer and a collimation of typically 30' downstream both of the monochromator and the sample. Very tight 10' collimators are also available.
The analyzer is composed of an array of four flat HOPG crystals. A well shielded single $^3$He counter is used for this option with a low background rate of less than one count per two minutes. Because MIRA is the first of several beamlines on NL6, it is operated in the constant-$k_i$ mode.

\begin{figure}[htb]
  \begin{center}
    \includegraphics[width=60mm]{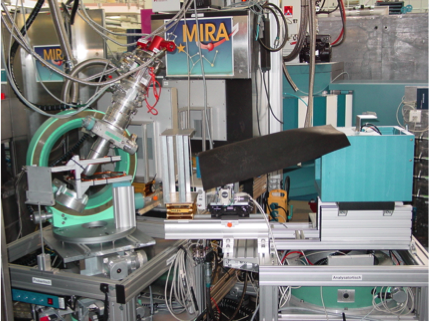}
    \caption{\label{fig:Pol_diff} Setup for cold neutrons using an Eulerian cradle and polarization analysis for the separation of the nuclear, magnetic, and chiral cross sections in magnetic materials.}
  \end{center}
\end{figure}

\subparagraph*{A cold (polarized) diffraction mode} with a high $Q$-resolution in the order of $10^{-3}$~\AA$^{-1}$. An Eulerian cradle (see Fig.~\ref{fig:Pol_diff}) gives access to a larger part of reciprocal space. Using an S-Bender with  $P= 98.5$ \% \cite{Krist} as polarizer and a polarizing V-cavity as analyzer the nuclear, magnetic, and chiral part of the scattered intensity can be separated. Further applications of the diffraction mode are i) the determination of the functionality of lipid membranes similarly as it is done at the V1 instrument at the BER-II reactor at HZB in Berlin \cite{Hauss_2016} and ii) the determination of the water content of bentonite sands during casting processes via the structural change of the first Bragg peak from the interstitial water layer \cite{Jordan_2013}.

\begin{figure}[htb]
  \begin{center}
    \includegraphics[width=80mm]{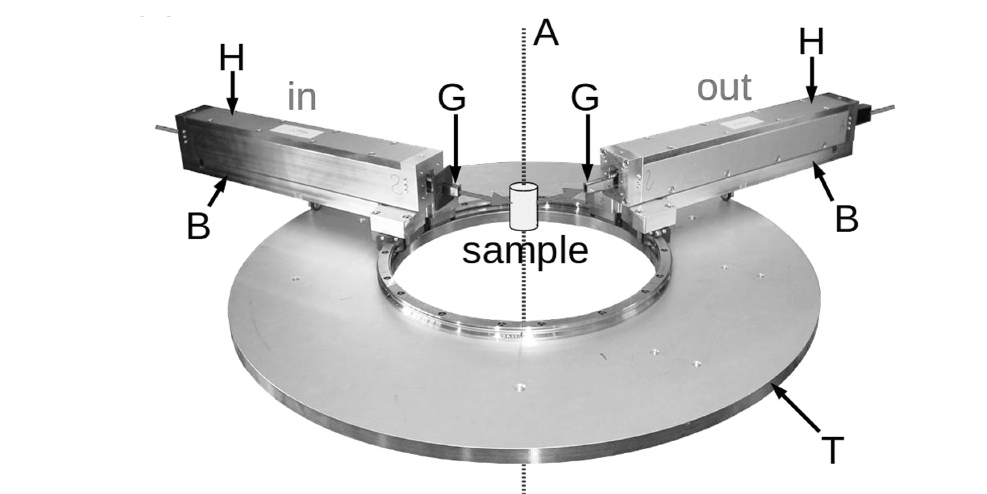}
    \includegraphics[width=50mm]{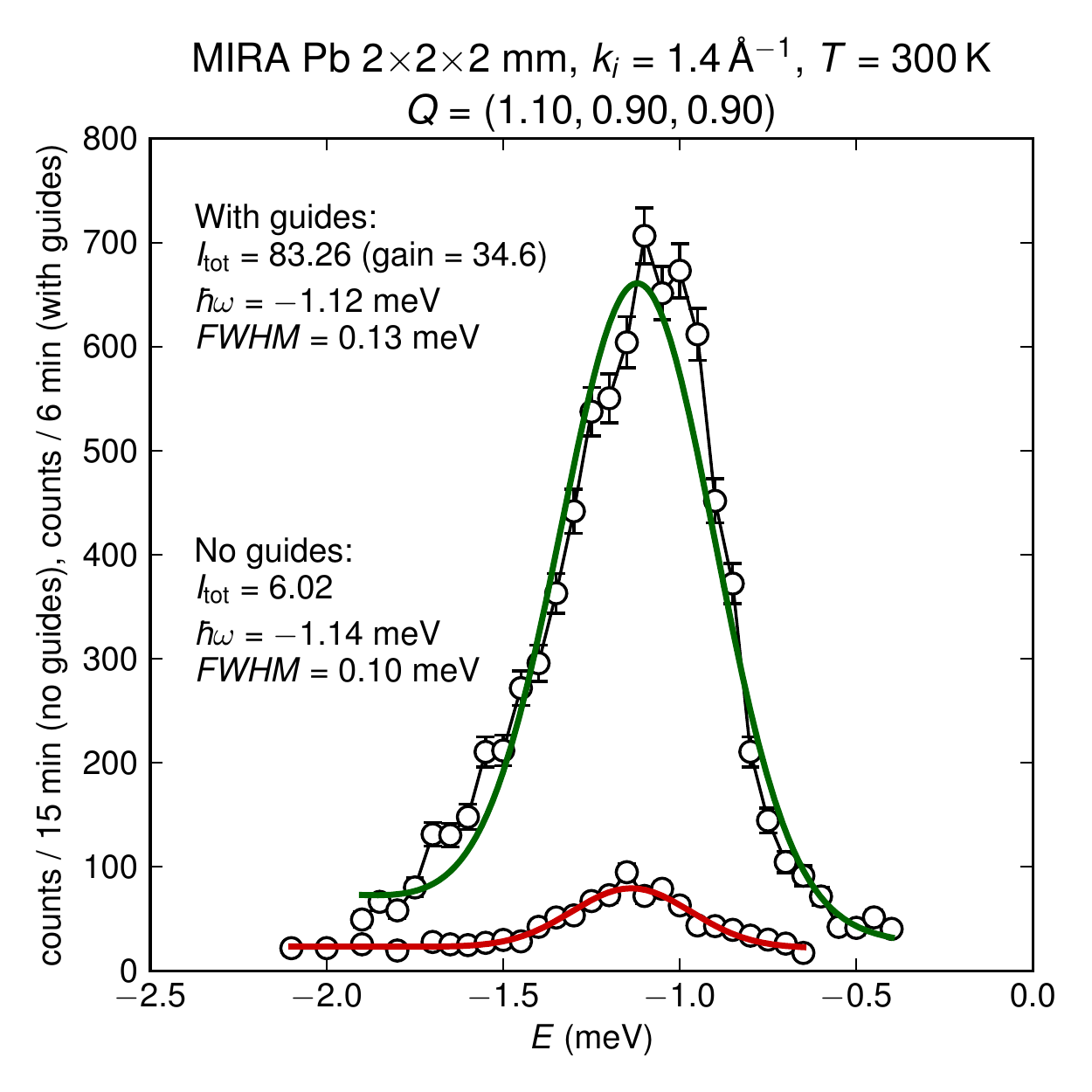}
    \caption{\label{fig:MIRA_layout} Top: The focusing guides are mounted on a rail system by means of kinematic mounts surrounding the sample.  Reprinted from \cite{Adams_2014}, with the permission of AIP Publishing. Bottom: Intensity gain due to the elliptic focusing guides. Comparison of a TA phonon as measured on a lead single crystal with a volume of 8 mm$^3$ without (red line) and with elliptic guides (green line). A gain factor of $G = 35$ is realized \cite{Brandl_2015}.}
  \end{center}
\end{figure}

\subparagraph*{A focusing mode using elliptic guides, suited for small samples and/or complex environments (pressure cells etc.)} for elastic and inelastic experiments. It utilizes polarizing and non-polarizing elliptic focusing guides with different focal lengths resulting in a very good signal-to-background ratio. The guides are mounted in housings that are equipped with kinematic mounts on a rail system thus allowing a quick exchange (Fig.~\ref{fig:MIRA_layout}, top). For more details see Refs. \cite{Adams_2014, Brandl_2015}.
There are limitations for the use of these focusing guides, which can be understood as follows: A homogeneous neutron beam at the entry of the guide, becomes inhomogeneous at the location of the sample, consisting of a rectangular pattern of nine maxima. The central maximum of this distribution is dominated by the direct beam, whereas the remaining eight maxima may be attributed to the sides and the corners of the guide. It is important to note that a second guide with its optical axis parallel to the first guide and with its focal point at the sample position, transforms the inhomogeneous phase space of the neutron beam at the sample location such that it is homogeneous again at the exit of the guides. When the optical axis of the focusing guide and the incident beam are {\it not} parallel, representing the most important example of a misalignment, the inhomogeneous phase space at the location of the sample is no longer symmetric. In turn, this may lead to an erroneous determination of the intensity of the diffraction peaks and the scattering angles \cite{Adams_2014}.

The biggest benefit of using focussing neutron guides may finally be expected, when the artefacts arising from the combination of inhomogeneous phase space and increased beam divergence at the sample location are not important. Prominent examples are inelastic neutron scattering studies, where large beam divergences are favorable. Here, major improvements of the signal to noise ratio exceeding well over an order of magnitude are obtained (see Fig.~\ref{fig:MIRA_layout}, bottom). This is especially useful for experiments using extreme conditions as high-pressure or high-magnetic fields.

\subparagraph*{A high-resolution spin-echo setup MIEZETOP} for measuring quasi-elastic dynamics in magnetic fields and from depolarizing samples is currently under construction. MIEZETOP is an add-on for MIRA based on the longitudinal neutron resonance spin-echo (LRNSE) technique \cite{Krautloher_2016}. It comprises all the polarizing equipment as polarizer, analyzer and guide fields as well as the precession region for the neutrons, similarly as the MIEZE setup at the beamline RESEDA \cite{RESEDA} at the FRM-II. In contrast to the conventional LNRSE technique, MIEZE requires only one arm for the precession of the neutrons yielding a time modulated signal that provides the intermediate scattering function $S(Q,\tau)$ \cite{Gaehler:92, Besenboeck:98, Georgii:11, Brandl:11}. The time dependence of the scattered neutrons is measured by means of a position sensitive GEM detector (CASCADE) covering an area $A = 20  \times 20$ cm$^2$ yielding a time resolution as small as picoseconds \cite{Klein:2011jj}. The whole MIEZETOP setup can easily be aligned with the help of an integrated auxiliary neutron guide thus allowing its use at other beamlines such as reflectometers and small angle neutron scattering (SANS).

\section{Scientific applications}
Typical applications of MIRA  are the determination of the structure and dynamics in single crystals exhibiting incommensurate magnetic order. Furthermore using the elliptic focal setup three axis measurements of small sample volumes can be performed. Some results obtained at MIRA are discussed in the following.

\subparagraph*{Diffraction from helimagnets and Skyrmion lattices.}
For these investigations, MIRA-1 is used in the SANS mode. Because of the available long wavelengths, SANS can be conducted even using this rather compact beamline.

\begin{figure}[h]
    \includegraphics[width=90mm]{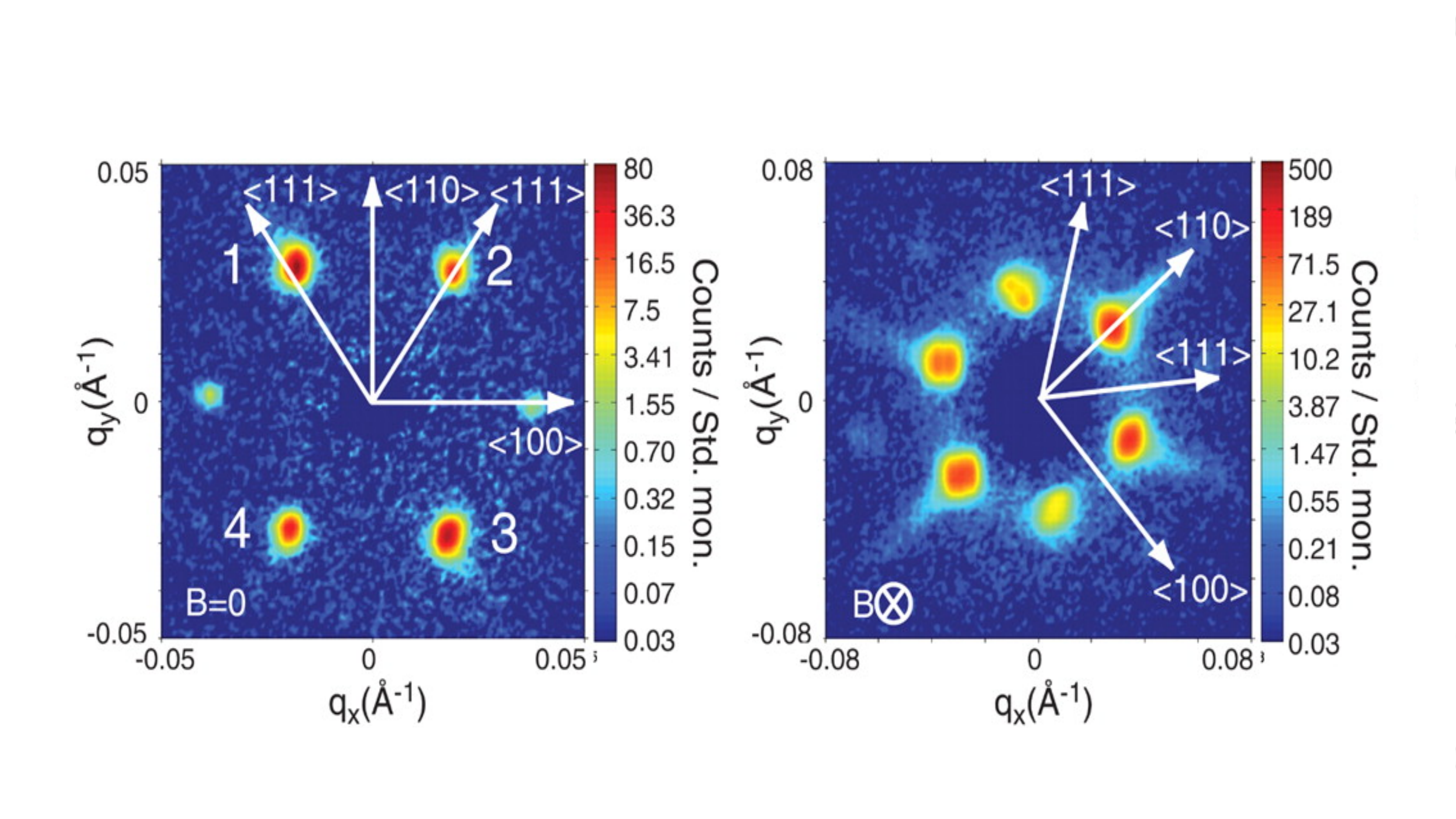}
      \begin{center}
    \includegraphics[width=70mm]{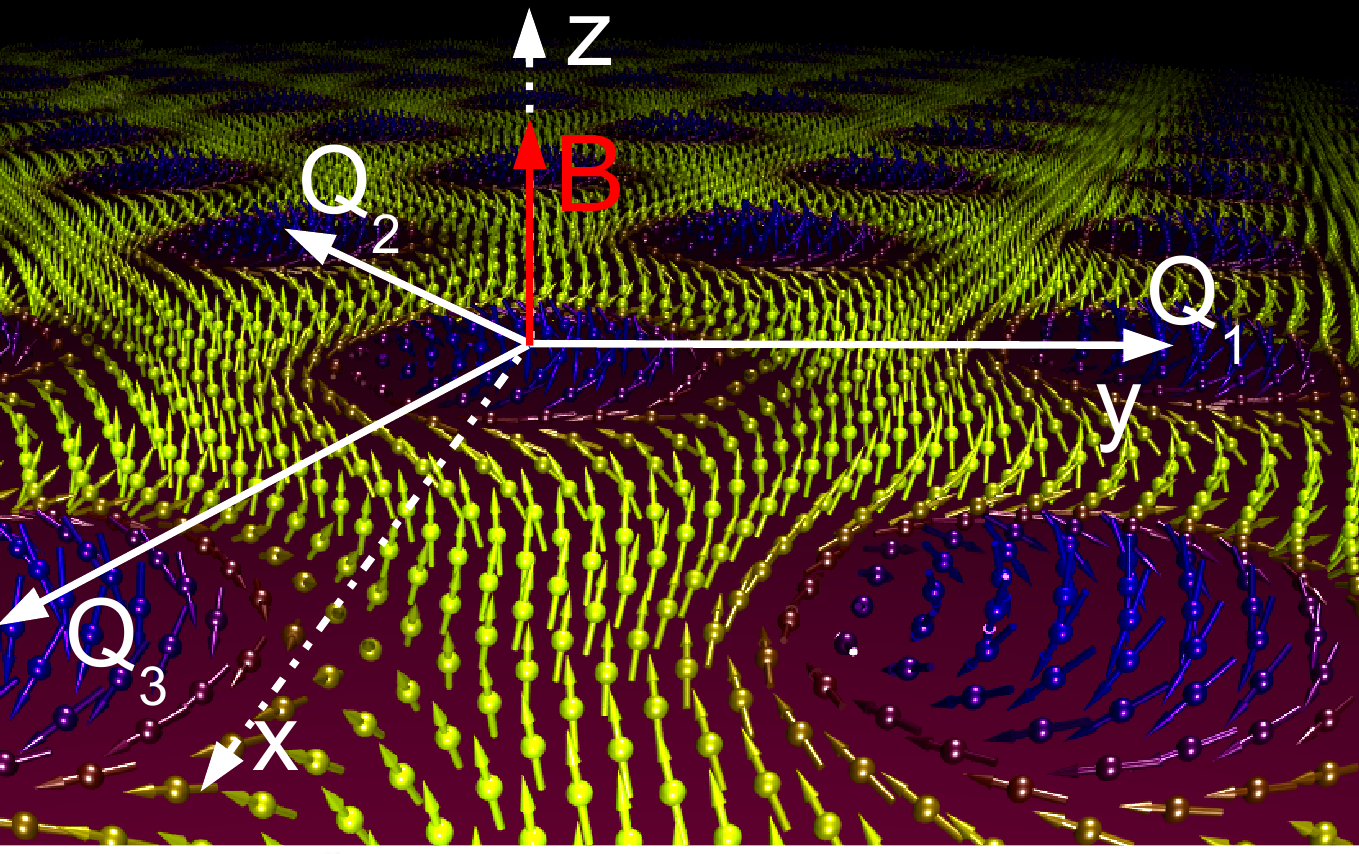}
    \caption{\label{fig:Skyrmions}Skyrmions in MnSi: (Top) The small angle scattering pattern of the 4-fold symmetry in the helical phase (left) contrary to the typical six-fold symmetry of  the Skyrmion lattice (right). Adopted  from \cite{SMuel:2009}. Reprinted with permission from AAAS. (Bottom) The real space picture of the skyrimion lattice.}
  \end{center}
\end{figure}

For example the prototype of chiral magnets MnSi has a magnetic structure, which is incommensurable with its long helimagnetic pitch of 175~\AA~compared to the lattice $d$-spacing of 4.2~\AA. Therefore the separation of the helix in small angle geometry with a $q$ of $3 \cdot 10^{-3} $~\AA~results in a $2\theta$-angle of 3 degrees for a wavelength of 10~\AA~(see Fig. \ref{fig:Skyrmions}, top left). 

Application of a field of 180 mT induces a Skyrmion lattice phase with its characteristic six-fold symmetry (Fig. \ref{fig:Skyrmions}, top right) \cite{SMuel:2009}. A real space visualization of the Skyrmion lattice is shown at the bottom of Fig. \ref{fig:Skyrmions}.

\subparagraph*{Helimagnons in MnSi.}
Goldstone modes in the helical phase of the weak itinerant ferromagnet MnSi have been determined for momentum transfers $\bf q$ parallel and 
perpendicular to the wave vector ${\bf k}_h$ of the helix using MIRA-2 in the TAS-mode \cite{Kugler_2015}. Because the helimagnetic state in MnSi is characterized by a long pitch of about 175\AA, the associated characteristic low-energy excitations can only be observed at very low reduced momentum transfer $q$. Therefore, the excellent $Q$ resolution available at MIRA was critical for this study. In order to overcome the problem of the superposition of excitations from the 4 chiral domains as experienced by Janoschek~\textit{et al}.~\cite{MJano:2010} we investigated the helimagnons in a single-$k_h$ state by applying a small magnetic field $B = 100$ mT. This way, the complexity of the magnon spectrum is significantly reduced allowing a detailed interpretation of the data with theory.

\begin{figure}[h]
  \begin{center}
    \includegraphics[width=70mm]{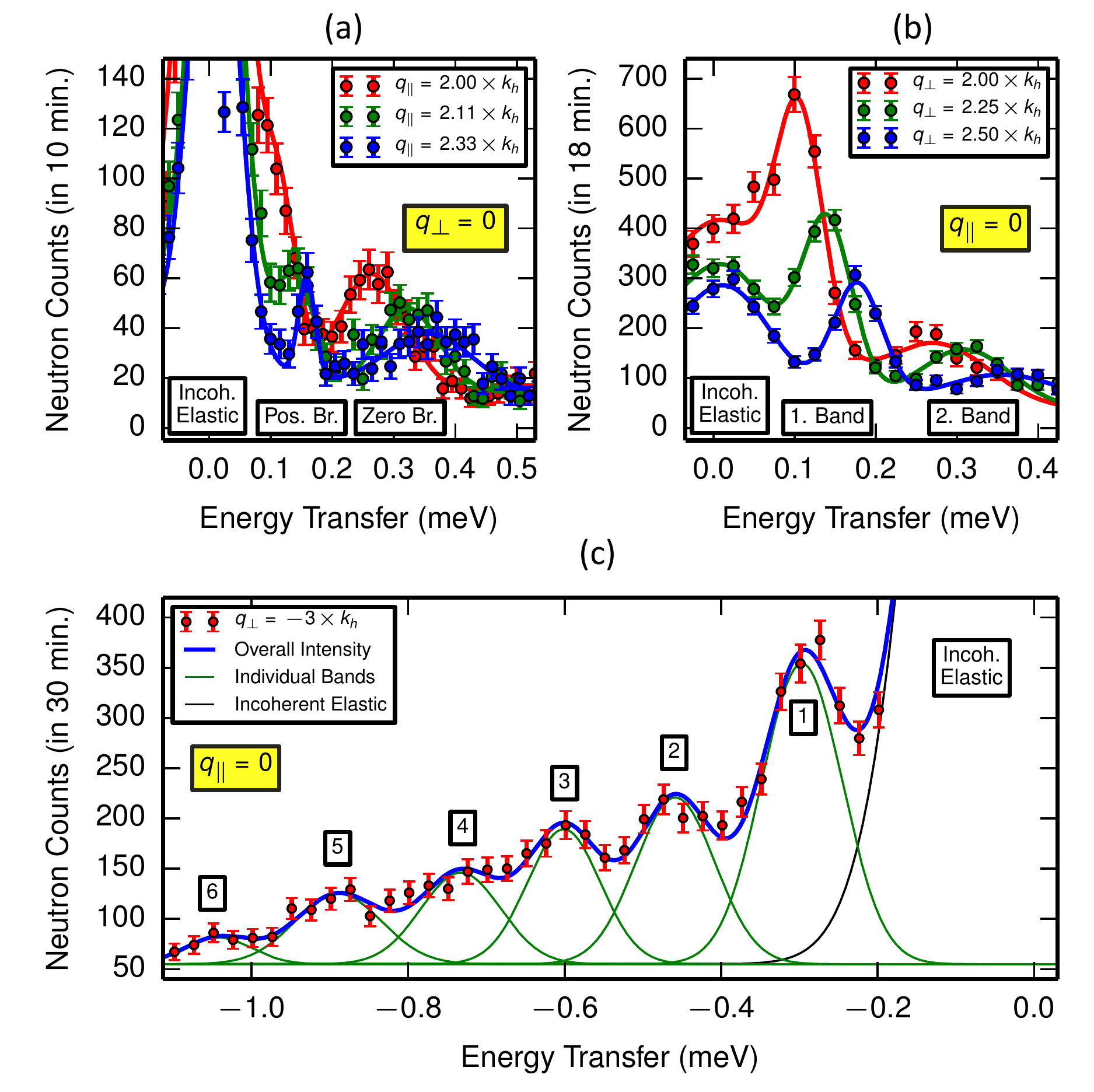}
    \caption{\label{fig:Helimagnons}Helimagnons in MnSi:  Shown are the three allowed helimagnon branches for ${\bf q \parallel k}_h$ (a) and the helimagnon bands which develop for ${\bf q \perp k}_h$ (b). Up to six bands can be resolved (c). Adapted with permission from Ref.~\cite{Kugler_2015}. Copyrighted by the American Physical Society.} 
  \end{center} 
 \end{figure}

The data is in perfect agreement with the parameter-free model for excitations in helimagnets by Belitz, Kirkpatrick, and Rosch \cite{Belitz_2006} when dipolar interactions and a higher order term are included. For $\bf q$ perpendicular to ${\bf k}_h$ a multiple band structure 
develops of which the first six bands can be resolved (Fig. \ref{fig:Helimagnons}). In addition, the renormalization of the helimagnons was investigated along and perpendicular to ${\bf k}_h$ demonstrating a very good agreement with magnetization data from the bulk.

\subparagraph*{Spindynamics in spinels.}
The usability and power of the focusing configuration was demonstrated during the course of the determination of the magnon spectrum in the spinel ZnCr$_2$Se$_4$ \cite{Brandl_2015}. Until now, the investigation of inelastic scattering was prohibitive because the available single crystals are extremely small.

\begin{figure}[htb]
  \begin{center}
    \includegraphics[width=75mm]{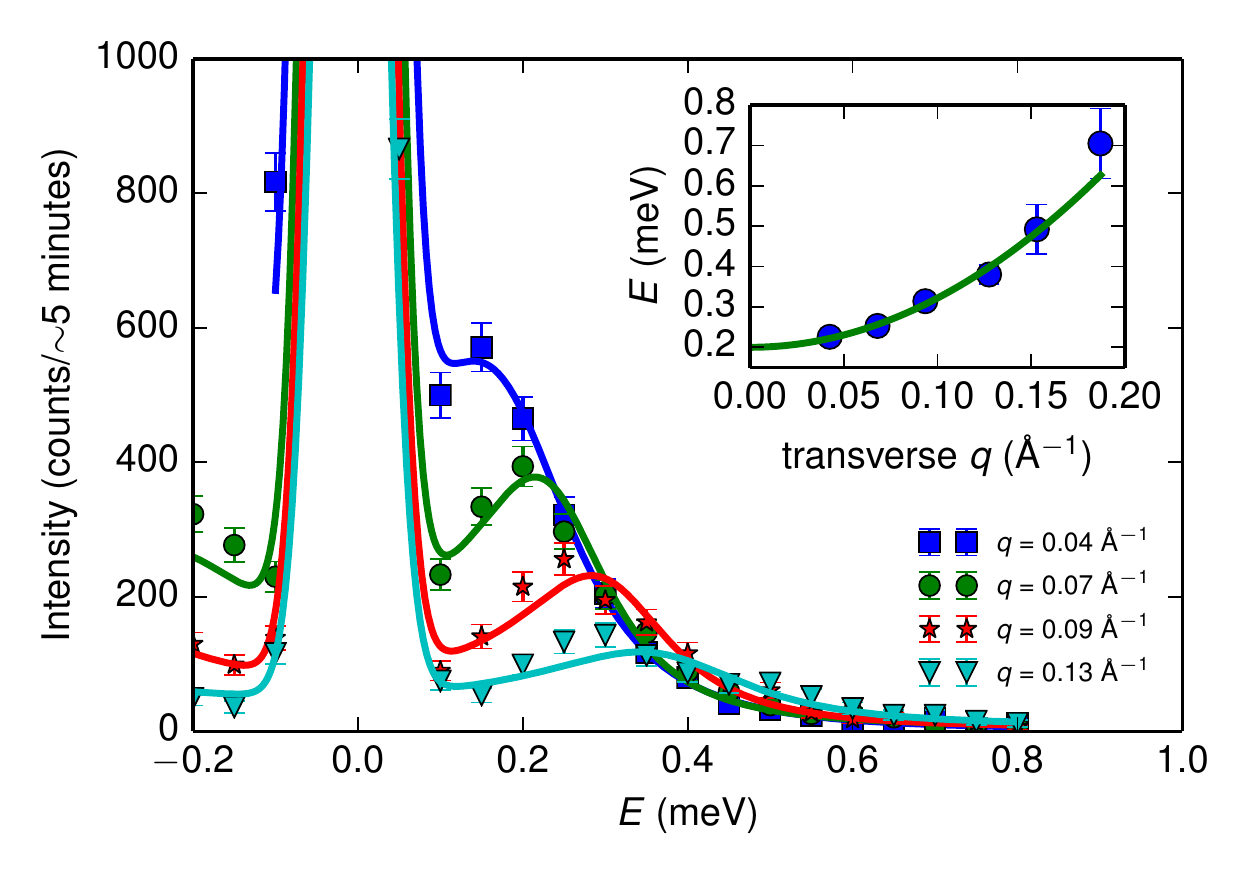}
    \caption{\label{fig:Spinel} Constant $Q$-scans in ZnCr$_2$Se$_4$. The energy increases with increasing momentum $q$. The dispersion as shown in the insert is quadratic in $q$. Reprinted from \cite{Brandl_2015}, with the permission of AIP Publishing.}
  \end{center}
\end{figure}

The magnon energies follow a quadratic dispersion relation $E_q=\Delta + D q^2$ as shown in the inset of Fig.~\ref{fig:Spinel} yielding an energy gap $\Delta$ = 0.2~meV and a stiffness $D = 8.8$~meV A$^2$. $D$ depends on the strength of the exchange interactions which are related to the magnetic ordering temperature. These results clearly demonstrate the dominance of the ferromagnetic exchange and provide the energy scale of the ferromagnetic coupling. Note that previous experiments without using focusing guides did not have sufficient neutron flux at the sample to distinguish any inelastic scattering from the instrumental background. 

\subparagraph*{Phase transitions in magnetic materials.} 
Two centuries of research on phase transitions have repeatedly highlighted the importance of critical fluctuations that show up in the vicinity of a critical point at $T = T_c$. If the phase space of the critical fluctuations becomes too large they may alter the order of the phase transition. The system may avoid the critical state altogether by undergoing a discontinuous first-order transition into an ordered phase. Fluctuation-induced first-order transitions have been discussed by Brazovskii \cite{Brazovskii} already many years ago. However, it was only recently that such a transition could clearly be identified at a magnetic phase transition. It was shown by Janoschek et al. \cite{Janoschek_2013}, that the phase transition from the paramagnetic to the helimagnetic state in MnSi can indeed be interpreted in terms of such a fluctuation induced phase transition.

\begin{figure}
  \begin{center}
    \includegraphics[width=70mm]{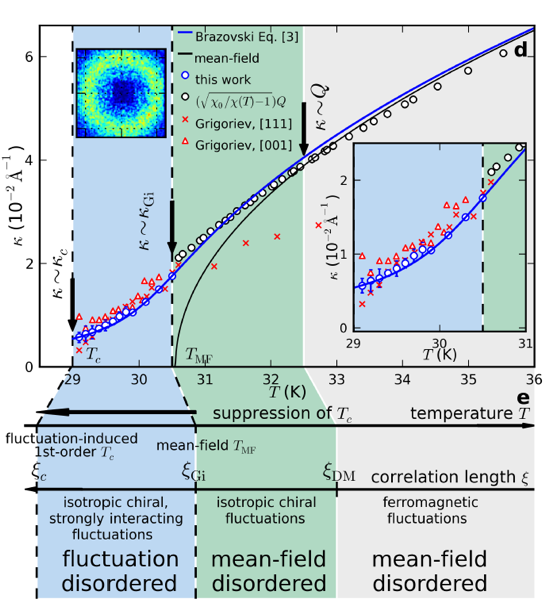}
    \caption{\label{fig:Bras} 
The figure shows the inverse magnetic correlation length $\kappa = 1/\xi$ versus $T$ (blue circles). The red symbols denote $\kappa$ according to previously published results \cite{Grigo_2005}. The blue solid line describes the renormalization of $\kappa$ as expected for a Brazovskii transition. The black solid line highlights the mean-field behavior well above $T_M$. The black circles are obtained from measurements of the magnetic susceptibility. The T-dependence of $\kappa$ indicates three separate regimes above $T_c$ namely a fluctuation disordered regime just above the fluctuation-induced first-order transition at $T_c$ and two mean-field disordered regimes. Adapted with permission from Ref.~\cite{Janoschek_2013}. Copyrighted by the American Physical Society.}
  \end{center}
\end{figure}

The SANS-experiments on MnSi using MIRA-1 show that the helical Bragg peaks exhibit a discontinuous jump at $T_c$ as expected for a first order phase transition. Above $T_c$ the magnetic intensity of the helical state is recovered entirely in the form of critical magnetic fluctuations on a sphere in momentum space. As shown in Fig.~\ref{fig:Bras}, the inverse magnetic correlation length $\kappa$ jumps at $T_c$ from zero to a finite value $\kappa_c$. This behavior is expected for a Brasovskii-type fluctuation induced first-order transition \cite{Janoschek_2013}. The temperature dependence of $\kappa$ can be nicely reproduced using the known parameters of the Dzyaloshinskii-Moriya helimagnet MnSi.


\section{Conclusions}

In this contribution we have highlighted some of the strengths of the multi-purpose three-axis spectrometer MIRA at the neutron source FRM II. The design of the beamline allows to perform elastic and inelastic neutron scattering experiments with high resolution in momentum ($Q$) and energy ($E$). Moreover, focusing neutron guides and MIEZETOP allow the investigation of small samples and measurements with very high $E$-resolution, respectively. Therefore, MIRA is particularly well suited for the investigation of excitations in materials with incommensurate order which require usually excellent resolution in $Q$ and $E$. In addition, MIRA provides well tailored neutron beams for investigations under extreme conditions.

\section{Acknowledgements}
We acknowledge technical support by Heinz Wagensonner, Reinhard Schwikowski, and Jonathan Frank, the design and construction of the neutron guides by the neutron optics group of FRM II, and the reliable and safe operation of the reactor by the reactor team of FRM II. MIEZETOP is funded by BMFT under the contract number 05K16WO4.
 
\section{References}
\bibliography{paper}
\end{document}